\documentclass[
preprintnumbers,
showpacs,
amsmath,
showkeys,
amssymb,
aps,
superscriptaddress,
twocolumn,
longbibliography,
letterpaper,
]{revtex4-2}
\usepackage{lipsum, babel}
\usepackage{color}
\usepackage{hyperref}
\usepackage{bm}
\usepackage{amsfonts}
\usepackage{mathrsfs}
\usepackage{graphicx}
\usepackage{amsmath}
\usepackage{float}
\usepackage{braket}
\usepackage{natbib}
\usepackage{amsmath}
\usepackage{color,graphicx,amsfonts,multirow}
\usepackage{booktabs}

\begin{document}
	
\title{Revisiting the deuteron mass radius via near-threshold $\rho^0$, $\omega$ and $\phi$ meson photoproduction}

\author{Xiaoxuan Lin}
\affiliation{Institute of Modern Physics, Chinese Academy of Sciences, Lanzhou 730000, China}
\affiliation{College of Physics Science and Technology, Hebei University, Baoding 71002, China}
	
\author{Wei Kou}
\affiliation{Institute of Modern Physics, Chinese Academy of Sciences, Lanzhou 730000, China}
\affiliation{School of Nuclear Science and Technology, University of Chinese Academy of Sciences, Beijing 100049, China}

\author{Shixin Fu}
\affiliation{Institute of Modern Physics, Chinese Academy of Sciences, Lanzhou 730000, China}
\affiliation{School of Nuclear Science and Technology, University of Chinese Academy of Sciences, Beijing 100049, China}

\author{Rong Wang}
\affiliation{Institute of Modern Physics, Chinese Academy of Sciences, Lanzhou 730000, China}
\affiliation{School of Nuclear Science and Technology, University of Chinese Academy of Sciences, Beijing 100049, China}
\affiliation{State Key Laboratory of Heavy Ion Science and Technology, Institute of Modern Physics, Chinese Academy of Sciences, Lanzhou 730000, China}

\author{Chengdong Han}
\email{chdhan@impcas.ac.cn (Corresponding Author)}
\affiliation{Institute of Modern Physics, Chinese Academy of Sciences, Lanzhou 730000, China}
\affiliation{School of Nuclear Science and Technology, University of Chinese Academy of Sciences, Beijing 100049, China}
\affiliation{State Key Laboratory of Heavy Ion Science and Technology, Institute of Modern Physics, Chinese Academy of Sciences, Lanzhou 730000, China}

\author{Xurong Chen}
\email{xchen@impcas.ac.cn (Corresponding Author)}
\affiliation{Institute of Modern Physics, Chinese Academy of Sciences, Lanzhou 730000, China}
\affiliation{School of Nuclear Science and Technology, University of Chinese Academy of Sciences, Beijing 100049, China}
\affiliation{State Key Laboratory of Heavy Ion Science and Technology, Institute of Modern Physics, Chinese Academy of Sciences, Lanzhou 730000, China}
\affiliation{Southern Center for Nuclear Science Theory, Institute of Modern Physics,Chinese Academy of Sciences, Huizhou 516000, China}

\begin{abstract}
We present a comprehensive analysis of near-threshold photoproduction of $\rho^0$, $\omega$, and $\phi$ mesons on a deuterium target, utilizing published datasets from DESY and SLAC for $\rho^0$ and $\omega$ production, as well as data from the LEPS and CLAS Collaborations for $\phi$ production. In extracting the deuteron mass radius, we adopt a dipole parametrization for the scalar gravitational form factor, which effectively captures the $|t|$-dependence of the differential cross sections associated with vector meson photoproduction. In addition, results from alternative commonly used form factor parametrizations are also considered and compared. Employing the vector meson dominance (VMD) framework and invoking low-energy Quantum Chromodynamics (QCD) theorems, we extract the deuteron mass radius from near-threshold photoproduction data of $\rho^0$, $\omega$, and $\phi$ mesons. The mass radii obtained from the various datasets are found to be consistent within statistical uncertainties, yielding an average value of $2.03 \pm 0.13$ fm under the dipole form assumption. We also provide a detailed discussion of the sensitivity of the extracted radius to different choices of gravitational form factor models. Our result represents a significant improvement in precision compared to earlier estimates based solely on $\phi$ meson photoproduction, offering new constraints for theoretical models of nuclear structure and deepening our understanding of the mass distribution within the deuteron.
\end{abstract}

	
\maketitle

\section{Introduction}
The size of proton, usually referred to as charge radius, magnetic radius, or mass radius, has always been a subject of heated discussions.
Due to the differences in proton charge radius observed in high-precision measurements,
the study of proton radius has always been the focal point of theoretical and experimental research,
which is often referred to as the proton charge radius puzzle~\cite{Pohl:2010zza, Antognini:2013txn, Carlson:2015jba}.
The nucleon magnetic radius \cite{Alarcon:2020kcz, Lin:2021umk, Djukanovic:2023jag} is a fundamental parameter that characterizes the spatial distribution of nucleon magnetization,
which originates from the motion and intrinsic magnetic moment of its constituent quarks and gluons.
The mass radius of nucleon describes the spatial distribution of mass within the nucleon, characterized by the mass density distribution.
As a fundamental property of composite systems, the mass radius spans a vast range of scales,
from subatomic particles in high-energy physics to galaxies in astrophysics.
Recently, significant progress has been made in the determination and interpretation of nucleon and light-nuclei mass radii
using various experimental and theoretical approaches \cite{Wang:2021ujy, Kharzeev:2021qkd, Wang:2021dis, Han:2022qet, Wang:2023uek}.
The mass of a particle can be regarded as the response of the particle to the external gravitational field,
and the gravitational form factor (GFF) of a particle is defined as the off-forward matrix element of the energy-momentum tensor (EMT)
in the particle state \cite{Kobzarev:1962wt, Pagels:1966zza, Teryaev:2016edw, Polyakov:2018zvc, Kharzeev:2021qkd}.
At low energy, the photoproduction of a quarkonium off the particle is connected to the scalar GFF of the particle, which is
sensitive to the particle mass distribution from the QCD trace anomaly.

Experimentally, the form factor $F(q)$ of the target is measured as a function of the momentum transfer $q$ in the low momentum elastic scattering process,
and represents the Fourier transform of the density distributions $\rho(r)$, which provides crucial insights into
the internal structure of the nucleon as described by QCD.
The form factor enables researcher to access information about internal energy-momentum
distributions inside the particle, thus directly linking experimental observables to fundamental QCD predictions~\cite{Ji:2021pys}.
For different hadronic systems, the internal density distribution is different,
which corresponds to different form factor parameterizations and root-mean-square radius of the particle.
Specifically, this mass radius provides insights into the mass density distribution within nuclear systems,
directly linked to underlying quark and gluon dynamics~\cite{Ji:1996ek, Burkert:2018bqq}.
The experimental determination of the proton and deuteron charge radii has also received renewed attention from facilities worldwide,
including experiments using electron-proton and electron-deuteron scattering, muonic atoms spectroscopy, and vector meson photoproduction
processes~\cite{Xiong:2019umf,Hammer:2019uab,Dupre:2020wop}.
Similarly, understanding the deuteron mass radius, which describes the mass distribution within the deuteron,
is essential for deepening our knowledge of the structure of the atomic nucleus.
Notably, recent analyses of vector meson photoproduction near production threshold have demonstrated
their capability as sensitive probes for extracting nuclear mass distributions and radii. 
Currently, based on some vector meson near-threshold photoproduction data, people have done several works on extracting the mass radii of
nucleon and light-nuclei \cite{Wang:2021ujy, Kharzeev:2021qkd, Wang:2021dis, Han:2022qet, Wang:2023uek}.

In this work, we systematically investigate the mass radius of the deuteron by analyzing the momentum transfer ($|t|$) dependence of differential cross sections
from near-threshold photoproduction of the vector mesons $\omega$, $\rho^0$, and $\phi$. By combining careful experimental data analysis with rigorous
theoretical modeling, we aim to provide reliable results contributing to the ongoing development of nuclear structure physics and QCD phenomenology. 
The organization of this paper is as follows: Section II briefly introduces the GFF and mass radius; Section III presents the data analysis and results; 
At the end, a short summary is given in Section IV.

\section{Gravitational form factor and mass radius}

In a nonrelativistic and weak gravitational field approximation, the scalar GFF provides a useful framework for describing a particle mass distribution.
That is, the mass radius of a particle can be theoretically defined in terms of the scalar GFF $G(t=q^{2})$, the form factor of the trace of the QCD EMT
instead of the form factor of $T_{00}$ \cite{Kharzeev:2021qkd}.
GFF can be obtained via the measurement of generalized parton distributions (GPDs) from various exclusive scattering processes,
as their second Mellin moments yield the combinations of GFFs \cite{Teryaev:2016edw, Polyakov:2018zvc}.
In this study, a possible approach is to transform the study of graviton-nucleon scattering into the scalar GFF of the nucleon
under the theoretical framework of the vector meson dominated (VMD) model.
For a continuous mass density distribution at small momentum transfer $t=q^2$,
the root-mean-square (RMS) radius of the nucleon is directly related to the slope of the scalar GFF at zero momentum transfer ($t=0$),
expressed by~\cite{Kumericki:2016ehc,Kharzeev:2021qkd,Miller:2018ybm},
\begin{equation}
  \left\langle R^2_m\right\rangle =6\frac{dG(t)}{dt}\Big|_{t=0},
\end{equation}
where the scalar GFF is normalized to $G(0)=M$ at zero momentum transfer $t=0$. 

For this analysis, the VMD model was used for describing near-threshold photoproduction processes of vector mesons on nuclear targets.
And the VMD model has been successfully applied in light vector meson near-threshold photoproductions
studies \cite{Strakovsky:2014wja, Strakovsky:2020uqs, Pentchev:2020kao, Wang:2022xpw, Han:2022khg, Han:2022btd}, accurately linking measurable
differential cross sections to the internal mass distributions of nucleon and light-nuclei~\cite{Wang:2021ujy, Kharzeev:2021qkd, Wang:2021dis, Han:2022qet, Wang:2023uek, Kou:2021bez},
as well as the vector-meson nucleon scattering lengths \cite{Strakovsky:2014wja, Strakovsky:2020uqs, Pentchev:2020kao, Wang:2022xpw, Han:2022khg, Han:2022btd}, etc.
Specifically, at energies near the photoproduction threshold and at small momentum transfer ($|t|\ll 1~\text{GeV}^2$), the VMD model approximates
the differential cross section by relating it directly to the square of the scalar GFF \cite{Wang:2021dis,Kharzeev:2021qkd}. Due to the near-nonrelativistic nature of heavy quarkonia produced in near-threshold photoproduction experiments, the coupling between the color-neutral heavy quark–antiquark pair and gluons can be described by the gluonic operator $g^2\mathbf{E}^{a2}$ \cite{Kharzeev:2021qkd}. This process is analogous to the Stark effect in QCD. The presence of this gluonic operator effectively encapsulates the interaction of the heavy quark pair with the gluon field in the nonrelativistic limit. Under these conditions, the near-threshold vector meson photoproduction amplitude can be factorized into a short-distance vacuum polarization contribution describing the photon splitting into a quark–antiquark pair, and a nonperturbative matrix element of the gluonic operator $g^2\mathbf{E}^{a2}$ evaluated between the initial and final nuclear states \cite{Kharzeev:2021qkd}
\begin{equation}
\mathcal{M}_{\gamma N\to\psi N}(t)=-e_qc_22M\langle p_2|g^2\mathbf{E}^{a2}|p_1\rangle,
\end{equation}
where $t=p_1-p_2$ is the momentum transfer, $e_q$ is the charge of quark and $c_2$ is a coefficient representing the short-distance coupling between a heavy quark and the color electric field. It also includes the process of the quark pair transforming into a vector meson.

At this stage, the gluonic operator can be expressed in terms of the trace part of the EMT. When inserted into the matrix element of the EMT, it yields a representation in terms of GFFs
\begin{equation}\mathcal{M}_{\gamma N\to\psi N}(t)=-e_qc_2\frac{16\pi^2M}{b_0}\langle p_2|T|p_1\rangle,\end{equation}
with
\begin{equation}\langle p_1|T|p_2\rangle=\left(\frac{M_N^2}{p_{01}p_{02}}\right)^{1/2}\bar{u}(p_1,s_1)u(p_2,s_2)G(t)\sim G(t).\end{equation}
Therefore, the differential cross section for quarkonium near-threshold photoproduction processes in the small $-|t|$ region can be described with the scalar GFF,
as shown below,
\begin{equation}
  \frac{d\sigma}{dt}\propto G^{2}(t).
  \label{eq:2}
\end{equation}
The validity of using the above equation to describe the photoproduction of light vector mesons off the nucleon or the hadronic matter is understandable. 
Brodsky $et\ al.$ \cite{Brodsky:1994kf} demonstrate that in the small momentum transfer regime, the distinctive features of forward differential cross section
for any possible vector meson leptoproduction can be reasonably factorized in perturbative QCD based on the $q\bar{q}$ wave function of the vector meson
and the target gluon distribution.

The dipole form factor from exponential distribution well describes the form factor of the nucleon within a wide kinematical range.
To quantitatively describe the $t$-dependence of the experimental data and subsequently extract the deuteron mass radius, we adopt a widely-used
dipole form parameterization for the scalar GFF~\cite{Kharzeev:2021qkd,Wang:2021dis},which is written as,
\begin{equation}
  G(t)=\frac{M}{\left(1-t/\Lambda^2\right)^2},
\end{equation}
where $\Lambda$ is dipole parameter determined from fitting differential cross-section data.
This simple and effective parameterization has demonstrated its ability to reliably reproduce experimental differential cross sections
across multiple vector meson photoproduction channels~\cite{Wang:2021dis}. More importantly, the dipole form factor has long been employed to describe double-gluon exchange processes in low-energy vector meson–nucleon scattering \cite{Frankfurt:2002ka}. This underpins the rationale that, within the weak-gravity approximation, the exchange of two gluons—or a tensor glueball—can effectively serve as a surrogate for graviton–nucleon or graviton–nucleus scattering. While it is practically impossible to construct a direct graviton–nucleon scattering experiment, such processes are holographically dual to each other in the framework of holographic QCD \cite{Mamo:2019mka,Mamo:2021krl,Mamo:2022eui}. Of course, the current experimental precision is not sufficient to rule out alternative form factor models. In what follows, we compare the commonly used parameterizations with the dipole ansatz.

\section{Data Analysis and Results}
Different hadronic systems exhibit distinct internal density distributions. Table   \ref{tab:model_forms_compare} summarizes several representative density profiles, along with their corresponding form factors and mean square radii, assuming $F(q)=G(q)/M$ is the normalized form factors. . For mesons, such as the pion, the density falls off rapidly with increasing radial distance and is typically described by a Yukawa-type potential. The associated form factor of the pion exhibits a monopole behavior. In contrast, the dipole form factor, derived from an exponential density distribution, provides a good description of the proton form factor over a broad range of momentum transfer. For heavy nuclei, such as lead, the density distribution reflects the saturation property of nuclear matter and can be approximately modeled by a uniform or Fermi distribution. Accordingly, we introduce several different parameterizations of GFFs to analyze the vector meson photoproduction data. In the latter part of this section, we first present the results of various form factor models in describing the data from different experimental collaborations, along with the corresponding extractions of the mass radius. A comprehensive analysis of the deuteron mass radius obtained from different models will be provided in the final subsection.
\begin{table}[htbp]
  \centering
  \begin{tabular}{lccc}
    \hline\hline
    \textbf{Model} & $\bm{\rho(r)}$ & $\bm{F(q)}$ & $\bm{\sqrt{\langle r^2 \rangle}}$ \\
    \hline
    \textbf{Yukawa-type}  & $\dfrac{\Lambda^2}{4\pi r} e^{-\Lambda r}$ & $\dfrac{1}{1 + q^2/\Lambda^2}$ & $\sqrt{\dfrac{6}{\Lambda^2}}$ \\
    \textbf{Exponential}  & $\dfrac{\Lambda^3}{8\pi} r e^{-\Lambda r}$ & $\dfrac{1}{(1 + q^2/\Lambda^2)^2}$ & $\sqrt{\dfrac{12}{\Lambda^2}}$ \\
    \textbf{Tripole}      & $\dfrac{\Lambda^4}{48\pi} r^2 e^{-\Lambda r}$ & $\dfrac{1}{(1 + q^2/\Lambda^2)^3}$ & $\sqrt{\dfrac{18}{\Lambda^2}}$ \\
    \textbf{Fourthpole}   & $\dfrac{\Lambda^5}{384\pi} r^3 e^{-\Lambda r}$ & $\dfrac{1}{(1 + q^2/\Lambda^2)^4}$ & $\sqrt{\dfrac{24}{\Lambda^2}}$ \\
    \textbf{Gaussian}     & $\left(\dfrac{\Lambda^2}{\pi}\right)^{3/2} e^{-\Lambda^2 r^2}$ & $e^{-q^2/(4\Lambda^2)}$ & $\sqrt{\dfrac{3}{2\Lambda^2}}$ \\
    \textbf{Uniform}      & $\dfrac{3}{4\pi R^3} \theta(R - r)$ & $\dfrac{3j_1(qR)}{qR}$ & $\sqrt{\dfrac{3R^2}{5}}$ \\
    \hline\hline
  \end{tabular}
   \caption{Analytical forms of density distributions $\rho(r)$, form factors $F(q)$, and corresponding root-mean-square (RMS) radii $\sqrt{\langle r^2 \rangle}$ for different models.}
   \label{tab:model_forms_compare}
\end{table}
\subsection{Photoproduction of $\rho^0$ on deuteron}

\begin{figure*}[htbp]
  \centering
  \includegraphics[width=0.7\textwidth]{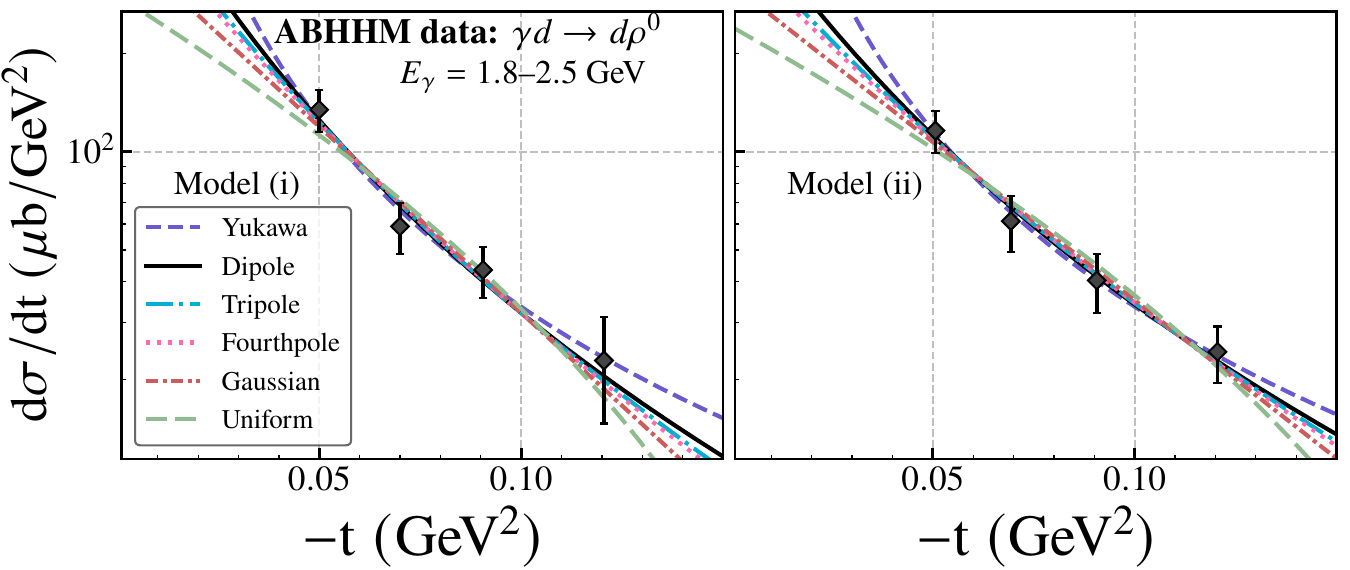}
  \caption{Differential cross sections of near-threshold $\rho^0$ photoproduction on deuteron fitted with six models using two resonance schemes (Model i and Model ii) from ABHHM Collaboration data~\cite{Benz1974}.}
  \label{fig:rho_6models}
\end{figure*}

The ABHHM collaboration~\cite{Benz1974} have been investigated the photoproduction of $\rho^0$ differential cross section in a deuterium
bubble chamber experiment at DESY with a bremsstrahlung beam at energies between 1 and 5 GeV.
We analyzed near-threshold $\rho^0$ meson photoproduction differential cross-section data on the deuterium target at $E_\gamma=1.8$-$2.5\,\text{GeV}$,
Two different resonance reconstruction models were considered, the Model (i) employs the standard Breit-Wigner resonance profile,
while Model (ii) incorporates the interference effects from Drell-type one-pion exchange.
Fig.~\ref{fig:rho_6models} shows the differential cross sections $d\sigma/dt$ of $\rho^0$ meson photonproductions
as a function of $-t$. In addition to the black solid curve corresponding to the dipole GFFs, the colored curves represent different parameterizations of the gravitational form factors. Evidently, in the low-$|t|$ region, the distinct shapes of these models—stemming from the lack of experimental constraints—highlight the critical importance of data in this kinematic domain.
Differential cross sections were independently fitted for each model to extract the corresponding deuteron mass radius, and from ABHHM data we extract the 
deuteron mass radii with different GFFs, respectively.
The extracted values of the parameter $\Lambda$ of GFFs and deuteron radius $\sqrt{\langle R^2_m\rangle}$ are listed in Table~\ref{table:rho_result}. It is evident that different models yield distinct descriptions of the same dataset, leading to variations in the extracted deuteron mass radius. However, the differences in the corresponding reduced $\chi^2$ values are not substantial. This indicates that, within the available kinematic range of the data, all considered GFFs parameterizations provide a reasonably good description of the differential cross sections.

\begin{table}[htbp]
	\centering
	\begin{tabular}{p{1.8cm}p{2.2cm}p{2.2cm}p{1.5cm}}
		\hline\hline
		\multicolumn{4}{c}{\textbf{ABHHM (Model i)}} \\
		\hline
		Model & $\Lambda$ (GeV) & $\sqrt{\langle R_m^2 \rangle}$ (fm) & $\chi^2/\text{ndf}$ \\
		Yukawa     & $0.06 \pm 0.02$ & $7.66 \pm 2.25$ & 0.85/2 \\
		Dipole     & $0.27 \pm 0.05$ & $2.52 \pm 0.47$ & 1.31/2 \\
		Tripole    & $0.38 \pm 0.05$ & $2.17 \pm 0.32$ & 1.55/2 \\
		Fourthpole & $0.48 \pm 0.06$ & $2.04 \pm 0.27$ & 1.68/2 \\
		Gaussian      & $0.14 \pm 0.01$ & $1.75 \pm 0.18$ & 2.21/2 \\
		\hline
		&$R\ (\mathrm{GeV}^{-1})$ &        &        \\
		Uniform    & $8.50 \pm 0.63$ & $1.30 \pm 0.10$ & 3.05/2 \\
		\hline
		\multicolumn{4}{c}{\textbf{ABHHM (Model ii)}} \\
		\hline
		Model & $\Lambda$ (GeV) & $\sqrt{\langle R_m^2 \rangle}$ (fm) & $\chi^2/\text{ndf}$ \\
		Yukawa     & $0.10 \pm 0.06$ & $5.31 \pm 3.85$ & 0.16/2 \\
		Dipole     & $0.31 \pm 0.05$ & $2.23 \pm 0.33$ & 0.56/2 \\
		Tripole    & $0.42 \pm 0.05$ & $1.96 \pm 0.23$ & 0.75/2 \\
		Fourthpole & $0.52 \pm 0.06$ & $1.85 \pm 0.20$ & 0.85/2 \\
		Gaussian      & $0.15 \pm 0.01$ & $1.61 \pm 0.14$ & 1.21/2 \\
		\hline
		&$R\ (\mathrm{GeV}^{-1})$ &        &        \\
		Uniform    & $8.00 \pm 0.50$ & $1.22 \pm 0.08$ & 1.95/2 \\
		\hline
		\multicolumn{4}{c}{\textbf{Global Average}} \\
		\hline
		Model & & $\sqrt{\langle R_m^2 \rangle}$ (fm) & ~ \\
		Yukawa     &  & $7.06 \pm 1.95$ & \\
		Dipole     &  & $2.33 \pm 0.27$ & \\
		Tripole    & & $2.03 \pm 0.18$ & \\
		Fourthpole & & $1.92 \pm 0.16$ & \\
		Gaussian      &  & $1.66 \pm 0.11$ & \\
		Uniform    & & $0.65 \pm 0.01$ & \\
		\hline\hline
	\end{tabular}
	\caption{Fitted results of $\Lambda$, deuteron mass radius $\sqrt{\langle R_m^2 \rangle}$, and $\chi^2/\text{ndf}$ for different models from near-threshold $\rho^0$ photoproduction data measured by the ABHHM Collaboration~\cite{Benz1974}. Global averages are calculated by combining results from Model i and ii.}
	\label{table:rho_result}
\end{table}

\subsection{Photoproduction of $\omega$ on deuteron}

\begin{figure}[htbp]
	\centering
	\includegraphics[width=0.49\textwidth]{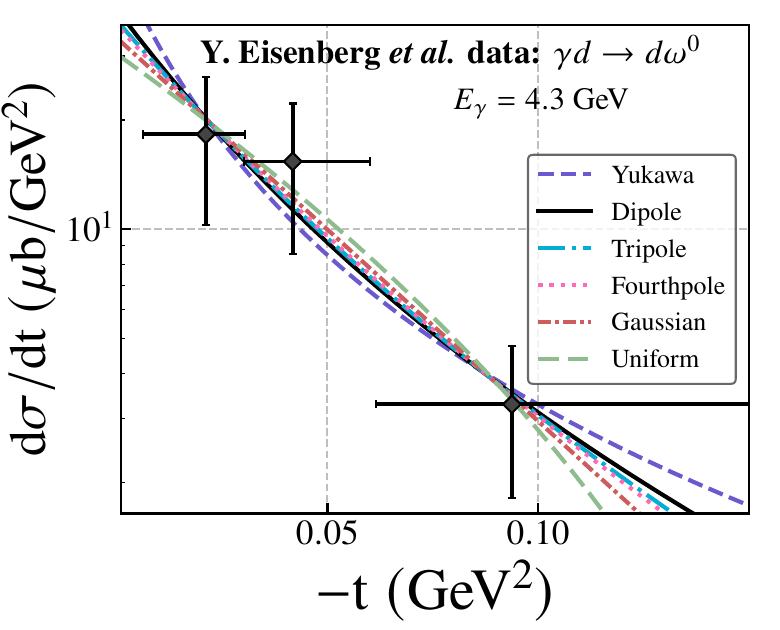}
	\caption{Differential cross sections of near-threshold $\omega$ photoproduction measured by Y.~\textit{Eisenberg et al.}~\cite{Eisenberg1972}, fitted with six models.}
	\label{fig:omega_6models}
\end{figure}

Y. Eisenberg $et$ $al.$ \cite{Eisenberg1972} have measured the coherent photoproduction of $\omega$ in $\gamma$d interaction at $E_\gamma$ = 4.3 GeV.
And the experiment was conducted by exposing the SLAC 40-inch bubble chamber to a quasi-monochromatic e$^{+}$ annihilation photon beam with an energy of 4.3 GeV. Fig.~\ref{fig:omega_6models} shows the differential cross sections of the near-threshold $\omega$ photoprodution reaction $\gamma$d $\rightarrow$ $\omega$d at $E_\gamma$ = 4.3 GeV.
From the fits of the dipole GFF, $\omega$ photoproduction differential cross-section data implies the deuteron mass radius to be $2.04\pm0.47\,\text{fm}$. 
The solid black curve show the fits of the dipole scalar GFF model, the colored curves represent the others models. 
To quantify the quality of fit, the obtained parameter $\Lambda$'s and the extracted deuteron radii $\sqrt{\langle R^2_m\rangle}$ 
from the differential cross sections of $\omega$ photoproduction near threshold are listed in Table~\ref{table:omega_result}. It can be observed that, in contrast to the case of $\rho^0$ meson production, the deuteron mass radii extracted from the $\omega$ production data do not exhibit significant variations across different GFF models. All extracted radii fall within the range of approximately 1 to 2~fm, and the corresponding reduced $\chi^2$ values remain relatively small. This behavior can be attributed to the fact that the $\omega$ production data lie within a lower $|t|$ kinematic regime, which provides somewhat stronger constraints on the models.

\begin{table}[htbp]
	\centering
	\begin{tabular}{p{1.6cm}p{2.2cm}p{2.2cm}p{1.2cm}}
		\hline\hline
		\multicolumn{4}{c}{\textbf{Y.~\textit{Eisenberg et al.}}} \\
		\hline
		Model & $\Lambda$ (GeV) & $\sqrt{\langle R_m^2 \rangle}$ (fm) & $\chi^2/\text{ndf}$ \\
		Yukawa     & $0.18 \pm 0.07$ & $2.68 \pm 1.04$ & 0.62/1 \\
		Dipole     & $0.33 \pm 0.08$ & $2.04 \pm 0.47$ & 0.43/1 \\
		Tripole    & $0.44 \pm 0.09$ & $1.90 \pm 0.38$ & 0.38/1 \\
		Fourthpole & $0.52 \pm 0.10$ & $1.84 \pm 0.35$ & 0.35/1 \\
		Gaussian      & $0.14 \pm 0.02$ & $1.68 \pm 0.27$ & 0.28/1 \\
		\hline
		&$R\ (\mathrm{GeV}^{-1})$ &        &        \\
		Uniform    & $8.76 \pm 1.05$ & $1.34 \pm 0.16$ & 0.19/1 \\
		\hline\hline
	\end{tabular}
	\caption{Fitted values of $\Lambda$, deuteron mass radius $\sqrt{\langle R_m^2 \rangle}$, and fit quality $\chi^2/\text{ndf}$ from the near-threshold $\omega$ photoproduction data measured by Y.~\textit{Eisenberg et al.}~\cite{Eisenberg1972}.}
	\label{table:omega_result}
\end{table}

\subsection{Photoproduction of $\phi$ on deuteron}

In the previous work~\cite{Wang:2021ujy}, we analyzed the coherent $\phi$-photoproduction differential cross sections on deuterium target
from CLAS and LEPS collaborations \cite{CLAS:2013jlg, CLAS:2007xhu,LEPS:2005hax,Chang:2007fc}, obtaining a deuteron mass radius of $1.95\pm0.19\,\text{fm}$.
This value is consistent with the values extracted from the current $\rho^0$ and $\omega$ meson analyses within the statistical uncertainties,
strongly supporting the robustness and universality of our theoretical approach across different vector meson channels. 

\begin{figure}[htbp]
	\centering
	\includegraphics[width=0.49\textwidth]{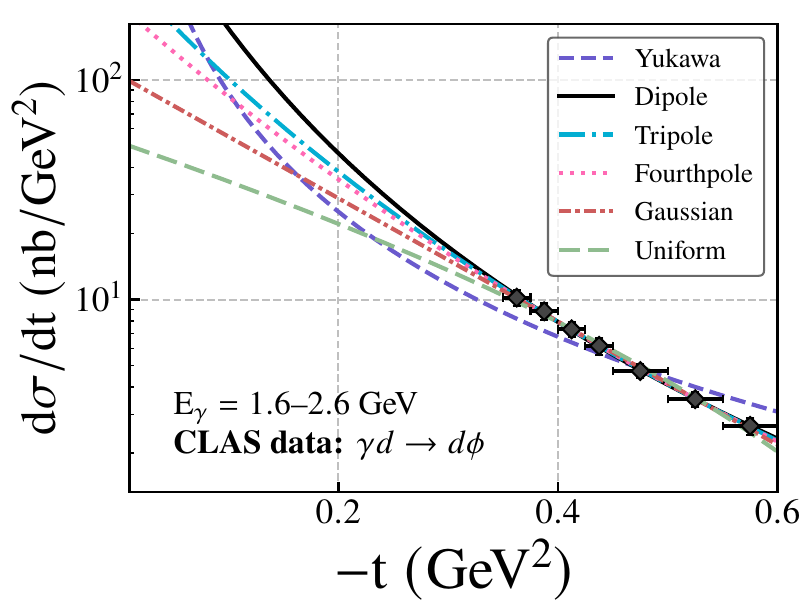}
	\caption{Differential cross sections of near-threshold $\phi$ photoproduction measured by CLAS~\cite{Mibe:2007aa}, fitted with six model parameterizations.}
	\label{fig:phi_clas_6models}
\end{figure}

\begin{figure*}[htbp]
	\centering
	\includegraphics[width=\textwidth]{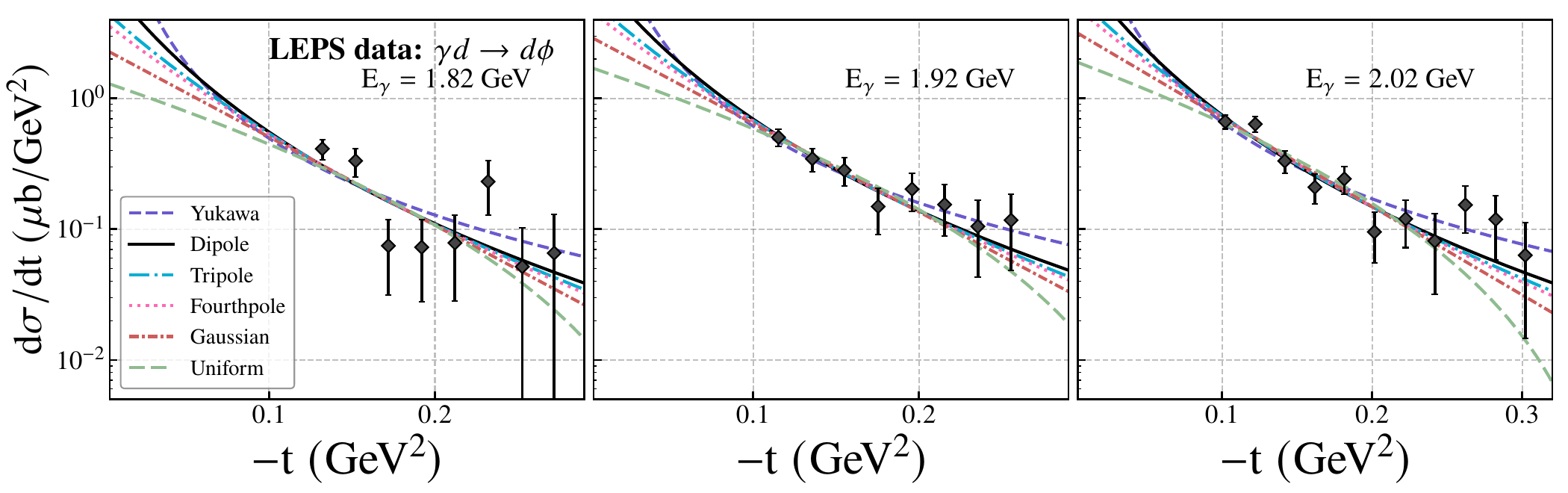}
	\caption{Differential cross sections of near-threshold $\phi$ photoproduction measured by the LEPS Collaboration~\cite{Mibe:2005er,Chang:2007fc}, fitted with six model parameterizations.}
	\label{fig:phi_leps_6models}
\end{figure*}

To enable a more detailed comparison, we applied the various form factor models to the differential cross-section data of $\phi$ photoproduction reported by the CLAS and LEPS collaborations. The corresponding fit results are shown in the Figs. \ref{fig:phi_clas_6models} and \ref{fig:phi_leps_6models}, and the extracted model parameters, including the deuteron mass radius, are summarized in the Table \ref{tab:radius_halfpage}. It is worth emphasizing that the LEPS collaboration provides data at three photon energy configurations, all of which can be regarded as near-threshold. We find that the results obtained by independently fitting the data at each energy point and subsequently averaging are consistent with those derived from a global fit using the combined dataset across all energy configurations. In principle, the deuteron mass radius is a fundamental property and should be independent of the specific photon energy configuration used in the experiment. Therefore, as long as the condition of near-threshold production is satisfied, the extracted results are expected to be largely insensitive to the precise photon energy. In addition, the CLAS collaboration data are predominantly located in the region of $|t|>0.3$ GeV$^2$, which leads to more pronounced variations in the fitted results. 


\begin{table}[htbp]
	\centering
	\begin{tabular}{p{1.6cm}p{2.2cm}p{2.2cm}p{1.2cm}}
		\hline\hline
		\multicolumn{4}{c}{\textbf{CLAS}} \\
		\hline
		Model & $\Lambda$ (GeV) & $\sqrt{\langle R_m^2 \rangle}$ (fm) & $\chi^2/\text{ndf}$ \\
		Yukawa     & $0.12 \pm 0.02$ & $3.86 \pm 0.76$ & 25.66/5 \\
		Dipole     & $0.40 \pm 0.05$ & $1.71 \pm 0.23$ & 0.20/5 \\
		Fourthpole & $0.88 \pm 0.05$ & $1.10 \pm 0.06$ & 0.23/5 \\
		Tripole    & $0.68 \pm 0.05$ & $1.22 \pm 0.08$ & 0.19/5 \\
		Gaussian     & $0.28 \pm 0.01$ & $0.87 \pm 0.03$ & 0.53/5 \\
		\hline
		&$R\ (\mathrm{GeV}^{-1})$ &        &        \\
		Uniform    & $3.99 \pm 0.08$ & $0.61 \pm 0.01$ & 1.93/5 \\
		
		\hline
		\multicolumn{4}{c}{\textbf{LEPS}} \\
		\hline
		Model & $\Lambda$ (GeV) & $\sqrt{\langle R_m^2 \rangle}$ (fm) & $\chi^2/\text{ndf}$ \\
		Yukawa     & $0.06 \pm 0.01$ & $8.26 \pm 1.45$ & 27.85/23 \\
		Dipole     & $0.32 \pm 0.04$ & $2.14 \pm 0.26$ & 24.01/23 \\
		Tripole    & $0.48 \pm 0.04$ & $1.75 \pm 0.15$ & 24.56/23 \\
		Fourthpole & $0.60 \pm 0.04$ & $1.62 \pm 0.12$ & 24.94/23 \\
		Gaussian      & $0.18 \pm 0.01$ & $1.35 \pm 0.07$ & 26.51/23 \\
		\hline
		&$R\ (\mathrm{GeV}^{-1})$ &        &        \\
		Uniform    & $6.32 \pm 0.24$ & $0.97 \pm 0.04$ & 31.83/23 \\
		\hline
		\multicolumn{4}{c}{\textbf{Global Average}} \\
		\hline
		Model &  & $\sqrt{\langle R_m^2 \rangle}$ (fm) & ~ \\
		Yukawa     &  & $4.82 \pm 0.68$ & \\
		Dipole     & & $1.90 \pm 0.17$ & \\
		Tripole    & & $1.36 \pm 0.07$ & \\
		Fourthpole & & $1.21 \pm 0.05$ & \\
		Gaussian      &  & $0.94 \pm 0.03$ & \\
		Uniform    && $0.65 \pm 0.01$ & \\
		\hline\hline
	\end{tabular}
	\caption{Fitted values of $\Lambda$, deuteron mass radius $\sqrt{\langle R_m^2 \rangle}$, and fit quality $\chi^2/\text{ndf}$ obtained from near-threshold $\phi$ photoproduction data measured by the CLAS Collaboration~\cite{Mibe:2007aa} and the LEPS Collaboration~\cite{Mibe:2005er,Chang:2007fc}.}
	\label{tab:radius_halfpage}
\end{table}

\subsection{Global analysis with $\rho^0$, $\omega$ and $\phi$ photoproduction on deuteron}

In summary, we have analyzed the near-threshold photoproduction data for all three types of vector mesons to extract the deuteron mass radius. The results, derived from different experimental collaborations, various vector meson probes, and distinct parameterizations of GFFs, consistently fall within the range of $1\sim 2$ fm. With the exception of the Yukawa-type GFFs---which yield a mass radius larger than the known deuteron charge radius---all other GFFs produce mass radii smaller than the charge radius \cite{CREMA:2016idx,Mohr:2024kco}. This indicates a strong possibility that the deuteron's energy density distribution is more compact than its charge distribution, although a certain degree of model dependence still persists at present.

\begin{figure*}[htbp]
	\centering
	\includegraphics[width=\textwidth]{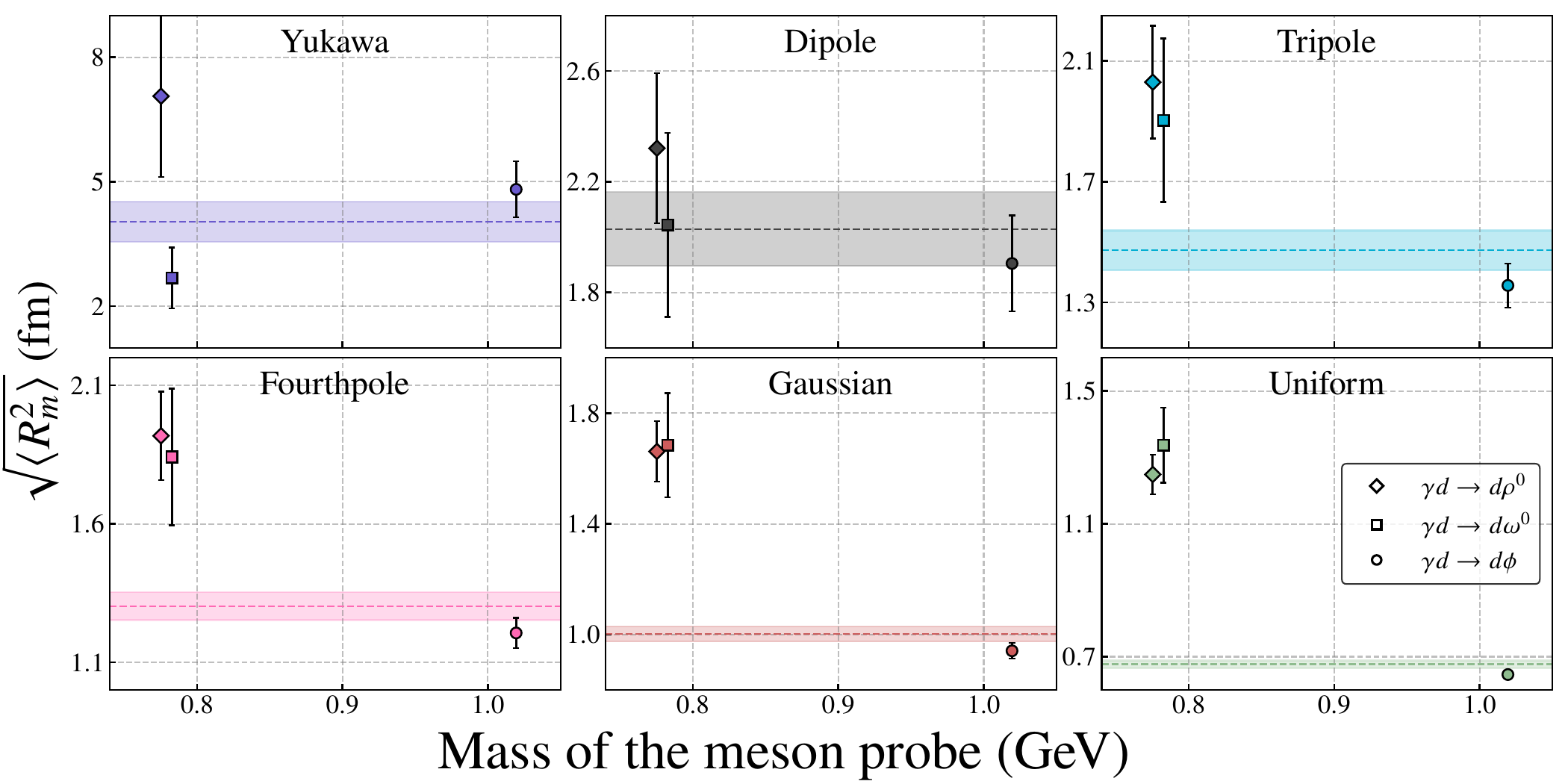}
	\caption{Comparison of deuteron mass radii extracted with different parameterizations as a function of meson probe mass. Each subplot corresponds to a specific model.}
	\label{fig:radius_vs_mass}
\end{figure*}

We used the following formula for the calculation of weighted average:
$\bar{x} \pm \delta \bar{x} = \sum_i w_i x_i/\sum_i w_i \pm \left( \sum_i w_i \right)^{-1/2} $ with $ w_i = 1/(\delta x_i)^2$.
Note that the weighted average of the mass radii with different models obtained here is consistent with the result of the simultaneous fit to all the data sets. We performed a weighted average of the deuteron mass radii obtained from each individual parameterization of GFFs. The weighted average results from all six GFFs parameterizations are presented in Fig. \ref{fig:radius_vs_mass}. The root-mean-square mass radii of the deuteron are extracted from the fits described above, where each GFFs model is used to describe the photoproduction data of $\rho^0$, $\omega$, and $\phi$ mesons. In each panel, the horizontal line and its associated uncertainty band represent the weighted average of the three data points. The resulting mass radii from the weighted averages corresponding to the six GFFs parameterizations are also summarized in the accompanying Table \ref{tab:final_radius_summary}.

\begin{table}[H]
	\centering
	\begin{tabular*}{0.35\textwidth}{@{\hskip 13pt}p{3.3cm}p{1.8cm}@{\hskip 2pt}} 
		\hline\hline
		Model & $\sqrt{\langle R_m^2 \rangle}$ (fm) \\
		\hline
		Yukawa      & $4.04 \pm 0.48$ \\
		Dipole      & $2.03 \pm 0.13$ \\
		Tripole     & $1.47 \pm 0.07$ \\
		Fourthpole  & $1.30 \pm 0.05$ \\
		Gaussian       & $1.00 \pm 0.03$ \\
		Uniform     & $0.68 \pm 0.01$ \\
		\hline\hline
	\end{tabular*}
	\caption{Final extracted deuteron mass radii $\sqrt{\langle R_m^2 \rangle}$ using different form factor models.}
		\label{tab:final_radius_summary}
\end{table}


From the deuteron mass radii extracted using six commonly adopted distribution models, it is evident that the results associated with $\rho^0$ and $\omega$ meson photoproduction exhibit larger uncertainties compared to those from $\phi$ meson data. Consequently, the weighted average is more heavily influenced by the $\phi$ meson results. Moreover, the dipole-type GFFs yield more consistent radii across all three vector meson channels, indicating a stronger generalization capability in describing near-threshold photoproduction processes. This suggests that dipole-type GFFs are more robust in modeling the underlying dynamics.
Future high-precision photoproduction experiments involving heavy vector mesons such as $J/\psi$ and $\Upsilon$ may provide additional constraints on GFFs. Based on these considerations, we favor extracting the deuteron mass radius within the framework of the two-gluon exchange picture, as encapsulated by dipole-type GFFs. This approach has already demonstrated its advantages in previous determinations of the proton and neutron mass radii.

\section{Summary}

Based on the assumptions of VMD model and a low energy QCD theorem, we extracted the deuteron mass radius by systematically analyzing different vector meson
near-threshold photoproduction data.
Utilizing a dipole parameterization of the scalar gravitational form factor within the VMD framework,
the combined analysis of the three vector mesons $\rho^0$, $\omega$ and $\phi$ photoproduction off the deuteron gives the average radius to be $2.03 \pm 0.13\,\text{fm}$ (dipole GFFs),
which is smaller than the world average of the deuteron charge radius (CODATA-2020 evaluation gives deuteron charge radius 2.1424 $\pm$ 0.0021 fm)  \cite{Mohr:2024kco,CREMA:2016idx}. 
This result is in good agreement with the previous determination based solely on $\phi$ photoproduction near-threshold data \cite{Wang:2021ujy},
thereby reinforcing the validity and universality of the adopted approach. Frankly speaking, our ultimate goal is to determine the deuteron mass radius in a manner that is independent of both the incident photon energy and the specific modeling choices. However, current experimental efforts toward extracting the mass radii of nucleons or nuclei still require extensive data accumulation and theoretical development. 
We argue that the deuteron mass radius, much like its charge radius, should be regarded as an intrinsic property that encodes information about the internal energy or mass distribution of the system. Nevertheless, our analysis indicates that different parameterizations of GFFs can yield notably distinct descriptions of the same set of differential cross-section data. This variation is primarily due to the limited number and precision of existing measurements, particularly the lack of data in the forward kinematic region, which significantly increases the uncertainties in the fits.
Taking all factors into consideration, we conclude that the dipole-type GFFs constitute one of the most reliable choices—both in terms of physical interpretability and fit quality. More stringent model constraints, however, will ultimately require high-precision measurements of forward differential cross sections.

This analysis provides not only necessary constraints for theoretical models of nuclear structure, but also a deeper understanding of the spatial distribution
of mass within the deuteron. The consistent deuteron mass extracted from different near-threshold vector meson photoproduction data indicates
that the VMD model combined with the dipole parameterization of the scalar GFF is a powerful tool to probe the internal structure of deuteron. 
In addition, precisely extracting the mass radius of deuteron is crucial for understanding the interplay between quark-gluon dynamics and nuclear binding,
as well as for refining our knowledge of generalized parton distributions, which are closely related to GFF.
Despite the success of our approach, several challenges remain. Currently, the approach to extracting the mass radius of deuterons parallels that of protons, meaning that the probe ``sees'' the structure of the entire nucleus without incorporating the shape of the deuteron itself into the size measurement process. Additionally, like the determination of the proton mass radius, different vector meson probes yield nearly identical deuteron mass radii within the uncertainty ranges, indicating that the form factor may not be sensitive to the type of meson probe used. The relatively large  uncertainties in some of the individual channels highlights the necessity
of improving experimental precision, especially in the low momentum-transfer region.
Future experimental campaigns at next-generation electron scattering facilities
and near-threshold photoproduction experiments are expected to provide high precision data, which will enable further improvement of the extraction of deuteron mass radius
and a more detailed mapping of the internal mass distribution.

In short, our comprehensive analysis in different vector meson photoproduction channels contributes to a deeper understanding of the deuteron structure
and lays a solid foundation for future theoretical and experimental investigations in nuclear mass radius physics. 
The  ingenious extraction of the deuteron mass radius from various meson photoproduction processes highlights the prospect of
combining advanced experimental techniques with rigorous theoretical modeling to reveal the complex structure of nuclear matter. 

\section*{Acknowledgments}
This work is supported by the National Natural Science Foundation of China under the Grant NO. 12305127,
the International Partnership Program of the Chinese Academy of Sciences under the Grant NO. 016GJHZ2022054FN.
and National Key R$\&$D Program of China under the Grant NO. 2024YFE0109800 and 2024YFE0109802.

\bibliography{refs}
\end{document}